\documentclass[pdflatex,sn-mathphys-num]{sn-jnl}

\usepackage{graphicx}%
\usepackage{multirow}%
\usepackage{amsmath,amssymb,amsfonts}%
\usepackage{amsthm}%
\usepackage{mathrsfs}%
\usepackage[title]{appendix}%
\usepackage{xcolor}%
\usepackage{textcomp}%
\usepackage{manyfoot}%
\usepackage{booktabs}%
\usepackage{algorithm}%
\usepackage{algorithmicx}%
\usepackage{algpseudocode}%
\usepackage{listings}%

\usepackage{mathtools}

\theoremstyle{thmstyleone}%
\theoremstyle{thmstyletwo}%

\theoremstyle{thmstylethree}%
\newcommand{\be}{\begin{equation}}
\newcommand{\ee}{\end{equation}}
\newcommand{\bea}{\begin{align}}
\newcommand{\bse}{\begin{subequations}}
\newcommand{\ese}{\end{subequations}}

\newcommand{\bvm}{|\vec{m}|}
\newcommand{\vm}{\vec{m}}
\newcommand{\vM}{\vec{M}}
\newcommand{\vh}{\vec{h}}
\newcommand{\vn}{\vec{n}}

\newcommand{\ti}{\tau_\text{i}}
\newcommand{\tg}{\widetilde\gamma}

\begin{document}

\title[Resonant spin mode from retarded relaxation]{Resonant spin mode from retarded relaxation}


\author{\fnm{Julia} \sur{Beater}}\email{julia.beater@tu-dortmund.de}

\author{\fnm{G\"otz S.} \sur{Uhrig}}\email{goetz.uhrig@tu-dortmund.de}

\affil{\orgdiv{Department of Physics}, \orgname{TU Dortmund University}, 
\orgaddress{\street{Otto-Hahn Stra\ss{}e~4}, \city{Dortmund}, \postcode{44227}, \state{NRW}, 
\country{Germany}}}


\abstract{In ultrafast spin dynamics additional modes in the THz range appear which cannot 
be captured by conventional Landau-Lifshitz equations. Extending these equations
by further time derivatives (inertial Landau-Lifshitz-Gilbert equation) 
solves the issue only partly. Here we suggest 
to explain these modes by retarded Lindbladian relaxation. In spite of the 
simplicity of this phenomenological approach the agreement with the
experimental data is promising: physically plausible parameters describe the
position and the large width of these resonances. The salient feature of
several almost equidistant resonance peaks is in-line with experimental evidence as well.}





\maketitle

\section{Introduction}\label{sec1}

Ultrafast magnetism is of particular current interest for fundamental reasons as well as for
the immense potential in applications in information technology 
\cite{chapp07,gomon14,chuma15,barma21,flebu24}. Typically, the unitary magnetic dynamics induced
by a Hermitian Hamiltonian can be accounted for, because the dispersion of the magnons is known
from measurements or from \textit{ab initio} calculations. But a second crucial ingredient
is the relaxation due to the interaction with environmental baths. 
The corresponding processes can be of phononic \cite{hartm25} or of electronic origin \cite{lenzi25}.

If the processes are not ultrafast, the Landau-Lifshitz (LL) equation \cite{landa35}
\be
\label{eq:ll}
\frac{d \vm }{dt} = \vm\times\vh_0 -\frac{\lambda}{\bvm} \vm\times(\vm\times\vh_0)
\ee
describes the relaxation phenomenologically very well. 
Here the Hamiltonian is given by $H=-\vh_0\cdot\vm$, $\vm$ being the magnetization, concretely the
 expectation value of a spin, and $\vh_0$ a fixed external magnetic field with $h_0=g\mu_\text{B} B$.
The Landau-Lifshitz-Gilbert (LLG) equation \cite{gilbe04} 
\be
\label{eq:llg}
\frac{d \vm }{dt} = \vm\times\vh_0 -\frac{\lambda}{\bvm} \vm\times\frac{d \vm }{dt} 
\ee
essentially describes the
same relaxation dynamics for small values of the dimensionless relaxation parameter $\lambda$.
The LLG version has the merit that it does not lead to unphysical instabilities for large values of
$\lambda$. Both equations keep the length of $\vm$ constant and treat the spin as classical vector.

If one is aiming at ultrafast dynamics in the picosecond range it is far less clear that
the LL or LLG equation provide a sufficient description. One striking observation consists
in the ultrafast modulation of the length of the magnetization
due to an optical pulse with slow recovery afterwards \cite{beaur96}.

Yet, recent extensions of the LL/LLG approach advocate another direction, namely the 
addition of another time derivative to the LLG equation augmenting it to the so-called
inertial LLG (iLLG) equation
\be
\label{eq:illg}
\frac{d \vm }{dt} = \vm\times\vh_0 -\frac{\lambda}{\bvm} \vm\times\left( \frac{d \vm }{dt}
+ \ti \frac{d^2 \vm }{dt^2} \right)
\ee
where $\ti$ represents an additional parameter of temporal dimension. This equation is called
``inertial'' because the second temporal derivative resembles the usual classical equation of motion
of a body with inertia. Note that this form again keeps the length of the magnetization vector
constant.


The iLLG has been advocated first on theoretical considerations. The phase space is extended by
distinguishing the magnetic moment and the related angular momentum  in 
statistical ensembles of elementary magnetic moments \cite{ciorn11,wegro12}. For slow dynamics the LLG
equation results while for fast dynamics, on a time scale shorter than $\ti$, the iLLG 
needs to be considered and a nutational additional resonance is predicted \cite{olive12,olive15}. 

Often, the inertial term is derived from an interaction of the local spins with itinerant electrons 
\cite{bhatt12,kikuc15,sayad16b}. The key observation is that the relaxation due to the 
electronic environment happens with some retardation.
The second time derivative ensues from a truncated Taylor expansion while spatial locality is assumed 
as justified approximation for homogeneous magnetizations. An alternative view on the role of 
electronic degrees of freedom also leads to a second time derivative, but with a sign
opposite to an inertial term \cite{fahnl11,fahnl11err}.
In an extension of the spin-electron model, the prefactors of Gilbert damping and inertia
even acquire a time dependence \cite{bajpa19}. Still, the origin of the inertia term is
the retardation of the relaxation effects and the resulting equations preserve the magnitude
of the magnetization.

An alternative conclusion is drawn also from spin-electron coupling, but with emphasis on the
spin-orbit coupling. In this view, without this relativistic effect of spin-orbit coupling,
no inertial behavior would  be observable \cite{thoni17,monda17,monda18,monda23b}.

Evidence is accumulating that the retarded relaxation is at the origin of inertial spin effects.
Besides the calculations for electronic baths mentioned above there are also approaches
based on the coupling of spins to a bosonic bath of phononic origin which also allow for the
derivation of an inertial term as subleading term of a Taylor expansion of the bath dynamics
\cite{quare24,hartm25}.

There are also alternative justifications for the inertial terms and, more generally,
additional characteristic frequencies. Surface effects of nanomagnets can also yield nutational
dynamics \cite{basta18}. Another route for magnetic inertia focusses on the optical excitation 
process via itinerant electrons \cite{reyes25}.

On the experimental side, results are still scarce.
A conjecture that inertia plays a role in 
spin switching was observed in antiferromagnets \cite{kimel09}. It is not due to
relaxation but rather due to the dynamics of the N\'eel vector on the two sublattices.
Also in the switching of the average spin orientation in quantum dots an inertial
behavior has been observed and described \cite{heist15,scher19}. Yet these
phenomena are not related to an inertial term in extended LLG equations.

A first experimental estimate for the temporal delay $\ti$ in the iLLG was obtained 
for cobalt films in the range $0.2$ to $0.4$ ps \cite{li15d}. Quite recently, experiments
on thin films of various ferromagnets revealed nutational resonances in the THz regime
\cite{neera21,unika22}. Interestingly, a similar observation was made by another group, 
but for a resonance at about $0.1$ THz \cite{de23,de24}. 

It is the goal of this article to shed light on the origin of the observed resonance and
on the applicability of the iLLG. We will start from a complementary viewpoint which
involves another energy scale related to relaxation changing the length of the magnetization,
i.e., a longitudinal relaxation in contrast to the so far almost exclusively considered
transversal relaxation.

\section{Results}

\subsection{Derivation of equations}

We recall that one way to derive the LL equation starts from the quantum Lindblad equation for 
a single spin in a magnetic field varying in time \cite{uhrig25}. 
If the direction of the field is $\vn(t)$, i.e.,
$\vh(t) = h(t) \vn(t)$ the differential equation reads
\bea
\frac{d\vm}{dt}(t)  &= \vm(t)\times\vh(t) + 2\gamma S (S\vn(t)-\vm(t))
\nonumber\\
&\qquad -\gamma S \vn(t)\times(\vn(t)\times\vm(t)) 
\label{eq:lindblad-fm}
\end{align}
where $S$ is the spin length, i.e., the maximum length $\vm$ can take and $\gamma$
is the relaxation rate at zero temperature in the underlying Lindblad equation.
The first term is the usual precession; the second term expresses that the magnetization
$\vm$ tends to take its maximum length and the final term describes the relaxing spirals
as long as $\vm$ is not oriented along the external field. All fields are taken at the same time
because it is assumed that the temporal variation  of the field direction $\vn(t)$ is much slower
than the dissipation itself which is the usual Markovian assumption. 
Thus, the dissipator adapts instantaneously to the magnetic field.

Equation \eqref{eq:lindblad-fm} does not yet resemble the LL equation.  But the LL equation is retrieved from
the above expression using a local mean-field theory in which the effective local field seen by
each spin is given by the truly external magnetic field $\vh_0$, which is static, and the exchange field
resulting from the coupling to the surrounding spins. Assuming a homogeneous ferromagnetic
magnetization it is given by $J\vm(t)$ and thus $\vh(t) =\vh_0 +J\vm(t)$ from which
$\vn(t)=\vh(t)/|\vh(t)|$ ensues. The internal exchange field
is generically much larger than the external field so that it is indicated to 
expand \eqref{eq:lindblad-fm} in orders of $\vh_0$ keeping only the leading order, yielding
\bea
\frac{d\vm}{dt} &= {\vm}\times \vh_0 + 2\gamma S(\frac{S}{\bvm}\vm-\vm)  
\nonumber\\
& \qquad - \lambda \frac{SC}{\bvm^2}\, \vm\times(\vm\times\vh_0)
\label{eq:analytic1}
\end{align}
with $C\coloneqq 2S/|\vm| -1$ and $\lambda \coloneqq {\hbar\gamma}/{J}$. The first and the last term
together exactly represent the LL equation with $\lambda$ being the dimensionless relaxation
parameter which stems from the ratio of the dissipation in the Lindblad equation to the
internal exchange coupling. Interestingly, the second term acts on the length of the magnetization
and does so with its own rate $\gamma$. Typically, the magnetization quickly converges
to its saturation length, here $S$ at zero temperature, while the orientational relaxation governed
by $\lambda h_0$ is much slower \cite{uhrig25}. Previously, it was noted that a Fermi's Golden rule
ansatz to spin dynamics yields LLG type equation plus additional terms \cite{cygor14} 
as is here the case as well.

In view of the majority of previous studies summarized in the Introduction, it suggests itself
to pass from Markovian to non-Markovian dynamics by taking a  retardation of the relaxation 
into account. We do so in a straightforward phenomenological way by replacing the orientation $\vn(t)$ in
\eqref{eq:lindblad-fm} by $\vn(t-\tau)$ with same positive delay $\tau$. Hence, instead of 
\eqref{eq:lindblad-fm}, we consider
\bea
\frac{d\vm}{dt}(t)  &= \vm(t)\times\vh(t) + 2\gamma S (S\vn(t-\tau)-\vm(t))
\nonumber\\
&\qquad -\gamma S \vn(t-\tau)\times(\vn(t-\tau)\times\vm(t)) .
\label{eq:lindblad-delay}
\end{align}
This equation can be derived in complete analogy to the derivation of \eqref{eq:lindblad-fm}
in Ref.~\cite{uhrig25} if one replaces the Lindblad jump operators $B(t)$ by $B(t-\tau)$.
The delay $\tau$ plays a similar, but not identical, role as the time constant $\ti$ in the iLLG. 
Passing as before to the leading order in the external magnetic field $\vh_0$ we arrive at
\bse
\label{eq:rlll}
\begin{align}
  \frac{d\vm(t)}{dt} &= \vm(t) \times \vh_0 \label{eq:first}
	\\
	& \ + 2\gamma S \left(S \frac{\vm(t-\tau)}{|\vm(t-\tau)|} - \vm(t)\right) \label{eq:second}
	\\
	& \ - \gamma S \frac{\vm(t-\tau) \times (\vm(t-\tau) \times \vm(t))}{|\vm(t-\tau)|^2} \label{eq:third}
	\\
  & \ - \frac{2\lambda S^2}{|\vec{m}(t-\tau)|^3}\vm(t-\tau) \times (\vm(t-\tau) \times \vh_0) \label{eq:fourth}
	\\
  & \ +\frac{\lambda S}{|\vec{m}(t-\tau)|^2} \vm(t-\tau) \times (\vm(t) \times \vh_0) . \label{eq:fifth}
\end{align}
\ese
The first line \eqref{eq:first} is the standard precession term. The second and third line, \eqref{eq:second}
and \eqref{eq:third}, describe dynamics on the energy scale set by the relaxation rate $\gamma$. For $\tau=0$,
the line \eqref{eq:second} yields the second term in \eqref{eq:analytic1} while the third line \eqref{eq:third}
vanishes. The fourth and fifth line, \eqref{eq:fourth} and \eqref{eq:fifth}, describe dynamics on the 
energy scale 
set by $\lambda h_0$ as does the standard relaxation term of the LL equation. For $\tau=0$, these two lines
combine to the third term in  \eqref{eq:analytic1}. Henceforth, we address this equation as 
retarded Lindblad-Landau-Lifshitz equation, retarded LLL equation for short.

\subsection{Comparison to experimental data}

Here we compare the time dependence as evaluated from the iLLG and from the retarded LLL equation
to the measure data in Refs.\ \cite{neera21,unika22}. Due to the small experimental displacements about
the equilibrium positions we linearize the differential equations for the computation of the linear
susceptibility. In the LLL equation we replace $\vm/S$ by $\vM/M_S$, i.e., we replace the spin expectation values relative to its maximum value $S$ by the magnetization
relative to its maximum value $M_S$ which clearly justified in ferromagnetic systems. 
We assume the systems to be at sufficiently low temperature so that the calculations 
can be performed at zero temperature.

\begin{figure}
    \centering
    \includegraphics[width=0.98\columnwidth]{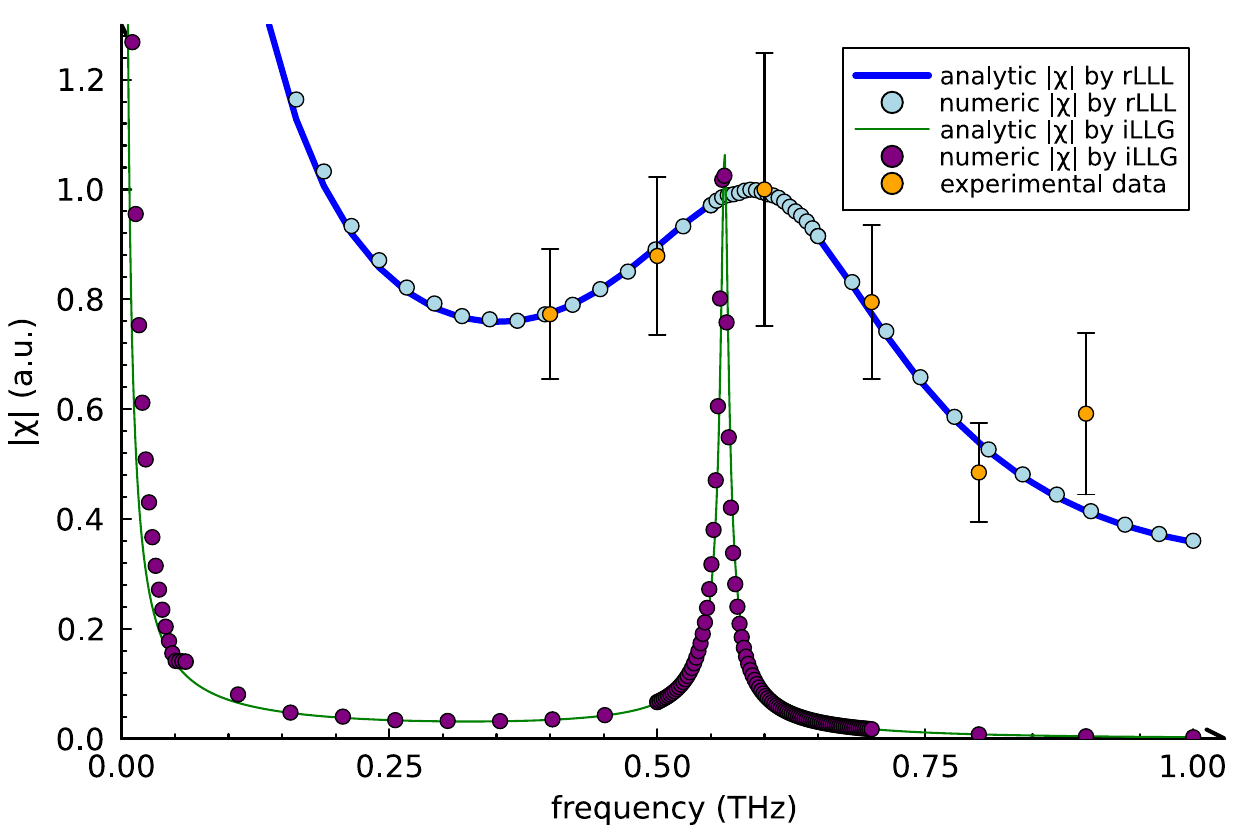}
    \caption{Dynamic susceptibility in arbitrary units 
		as measured in epitaxial permalloy in Ref.\ \cite{neera21} (yellow circles); as fitted with the 
		iLLG equation \eqref{eq:illg} with $\lambda=0.0058$ and $\ti=49$ps 
		(parameters from Ref.~\cite{neera21})
		from analytics (green line) and numerics (purple circles); 
		as fitted with retarded LLL equation \eqref{eq:rlll} with $\lambda=0.0058$, $\tau=1.26$\,ps, and 
		$\tilde\gamma = 1.28$\,ps$^{-1}$.}
    \label{fig:comparison}
\end{figure}

Figure \ref{fig:comparison} shows the instructive comparison for epitaxial permalloy 
(Ni$_{0.81}$Fe$_{0.19}$). 
The damping parameter $\lambda$ is taken from independent measurements in Ref.~\cite{neera21}
where the letter $\alpha$ was used for it. The time constants or retardation $\ti$ in the iLLG equation and 
$\tau$ in the retarded LLL are determined such that the resonance frequency matches.
The LLL equation requires and allows for an additional relaxation rate $\gamma$ which we discuss below.
For analyzing the measured curves quantitatively, it is convenient to use $\tg \coloneqq \gamma S$.

The first striking difference is the width of the resonance peak. Obviously, the Gilbert damping parameter
$\lambda$ in the iLLG equation does not capture the width adequately. This was already noted in the experimental 
analysis where the damping parameter was artificially increased by a factor 10. We take this observation 
as a first piece of evidence that the physics takes place on a higher energy scale.

Second, the required values of $\ti\approx 49$\,ps is too large in the following sense.
Mostly, the iLLG equations are derived from a Taylor expansion taking a retardation into account.
For this justification of the iLLG to make sense the additional second derivative should be smaller or
at worst of similar size than the leading first order term in \eqref{eq:illg}. 
Considering, however, the resonance at $0.56$\,THz the second term is larger than
the first term by the factor $2\pi\cdot 0.56\cdot 49\approx 170$, i.e., by two orders of magnitude too large for a justification by a Taylor expansion. At least for these experimental resonance the iLLG 
approach would require a derivation not using a Taylor expansion. 

Turning now to the retarded LLL approach with its additional parameter $\tg=1,28$\,ps$^{-1}$ 
which achieves an  excellent agreement with the experimental data. We note that the
retardation $\tau$ necessary to capture the resonance is in the range of picoseconds. This appears 
reasonable for a resonance in the THz range. Arguably, an additional fit parameter, here
$\tg$ helps to reach a good agreement. So, let us estimate its value to see whether it is in the
correct ballpark. We recall that $\lambda=\hbar\gamma/J$ so that the value of $\gamma$ and whence $\tg$
can be estimated if one knows $\lambda$ and the summarized exchange constant $J$. 
We use again  local mean-field theory which for collinear magnetism is the
same for the Heisenberg model and the Ising model \cite{strec15} and thereby relate $J$ 
to the Curie temperature $T_c$ via $J=3k_\text{B}T_c/(S(S+1))$.

This implies $\hbar\tg=3\lambda k_\text{B}T_c/(S+1)$. Using $T_c=815$\,K \cite{dijit18,zhang19c} 
and $S=0.46$ (see Methods)
we obtain $\tg_\text{estimate} \approx 1.27$\,ps$^{-1}$ in agreement with the $\tg$ from the fit.
By no means, we claim that the agreement is as good as these two numbers suggest, for instance in view
of the simple local mean-field formula used, see App.\ \ref{app:neera}
for the analysis of two more data sets from Ref.\ \cite{neera21}. 
But it clearly shows that the energy scale is
the correct one of the order of a fraction of the exchange field. Hence this energy scale
suggests itself as the high energy scale 
responsible for the observed THz resonance. Note that this line of thought
also re-opens the question whether the resonance stems from a nutation at all.

\subsection{Analytical understanding}

In view of the complexity of Eq.\ \eqref{eq:rlll}, it is insightful to strip off the
less important parts and to focus on the essential part only. Two observations are important:
(i) the dynamics is an ultrafast one, rather at the scale $\tg$ than at the scale $\lambda h_0$;
(ii) the resonance is fairly broad with a ratio of FWHM to resonance frequency of about unity.
How does this result from Eq.\ \eqref{eq:rlll}?

After passing from $\vm$ to $\vM$ and leaving out the slow terms $\propto \lambda$
or $\propto h_0$ we stay 
only with lines \eqref{eq:second} and \eqref{eq:third} in \eqref{eq:rlll} yielding
\bse
\label{eq:he}
\begin{align}
 \frac{d\vM(t)}{dt} &= 2\gamma S \left(M_S \frac{\vM(t-\tau)}{|\vM(t-\tau)|} - \vM(t)\right) \label{eq:he-first}
	\\
	& \ - \gamma S \frac{\vM(t-\tau) \times (\vM(t-\tau) \times \vM(t))}{|\vM(t-\tau)|^2} \label{eq:he-second}.
\end{align}
\ese
Focussing on small deviations from the equilibrium saturated orientation as they are induced in 
experiment we can distinguish longitudinal deviations $\vM=\vM_S + \vM_\parallel$ and
transversal deviations $\vM=\vM_S + \vM_\perp$ where $\parallel$ and $\perp$ are taken
relative to $\vM_S$. In the longitudinal case, the second line \eqref{eq:he-second} vanishes and 
from the first line only 
\be
 \frac{d\vM_\parallel(t)}{dt} = - 2\tg \vM_\parallel(t)
\ee
persists. This clearly describes a quick convergence to the equilibrium value $\vM_\parallel=0$;
it does not support an oscillating solution.

In the transversal case, line 1 and line 2 in \eqref{eq:he} partly compensate and
the remaining equation reads
\be
\frac{d\vM_\perp(t)}{dt} = \tg \left(\vM_\perp(t-\tau) - \vM_\perp(t)\right) .
\ee
This is an intriguing difference equation which is diagonal in the two perpendicular directions.
Thus, we can omit the vector character and use the ansatz $M(t) =\exp(-i\omega t) M(\omega)$
yielding the algebraic equation for the eigen modes $-i\omega = \tg (\exp(i\omega\tau)-1)$.
In order to deal with dimensionless parameters we multiply by $\tau$ and introduce
$x=\omega\tau$ and $\alpha=\tg\tau$ to obtain
\be
\label{eq:algebra}
-ix = \alpha(\exp(ix) -1).
\ee 
For given parameter $\alpha$ the solution $x$ describes the eigen mode. The
real part of $x$ defines the oscillatory circular frequency while its negative imaginary
part defines its decay rate. The existence of such solutions already answers key question (i), 
namely that oscillatory solutions exists, though these are not nutational.

We stress that these frequencies and decay rates are closely linked since there is
only one free parameter, i.e., they depend on the same parameter $\alpha$ and are not
independently tunable. This answers key question (ii), namely that the width
of the resonances is of the same order of magnitude than the resonance frequency.

\begin{figure}
    \centering
    \includegraphics[width=0.98\columnwidth,clip]{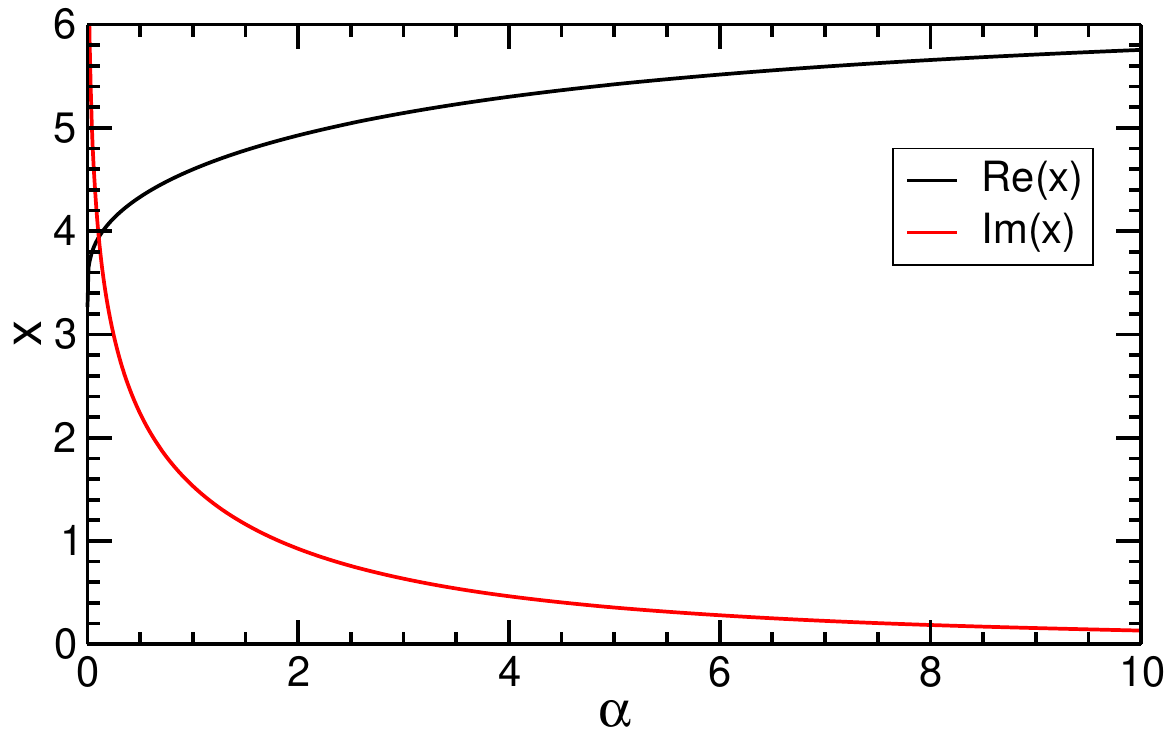}
    \caption{First solution $x$ of Eq.~\eqref{eq:algebra}
		as function of the parameter $\alpha$. The real part defines the oscillabory 
		circular frequency $\text{Re}(x)/\tau$ while the negative imaginary part determines
		the decay rate $-\text{Im}(x)/\tau$. The relative width of the resonances
		decreases for increasing $\alpha$.}
    \label{fig:quant-a}
\end{figure}

Quantitatively, we plot in Fig.~\ref{fig:quant-a} the frequency (real part of $x$) and the decay rate
 (negative imaginary part) of the resonance. 
Clearly, the resonances become
sharper for larger values of $\alpha$. But the ratios between decay and frequency are
of the order of unity unless $\alpha$ takes extreme values.
For this data we analyzed the first solution of $x$, but  it turns out that Eq.\ \eqref{eq:algebra}
allows for many solutions. 

\begin{figure}
    \centering
    		\includegraphics[width=0.98\columnwidth,clip]{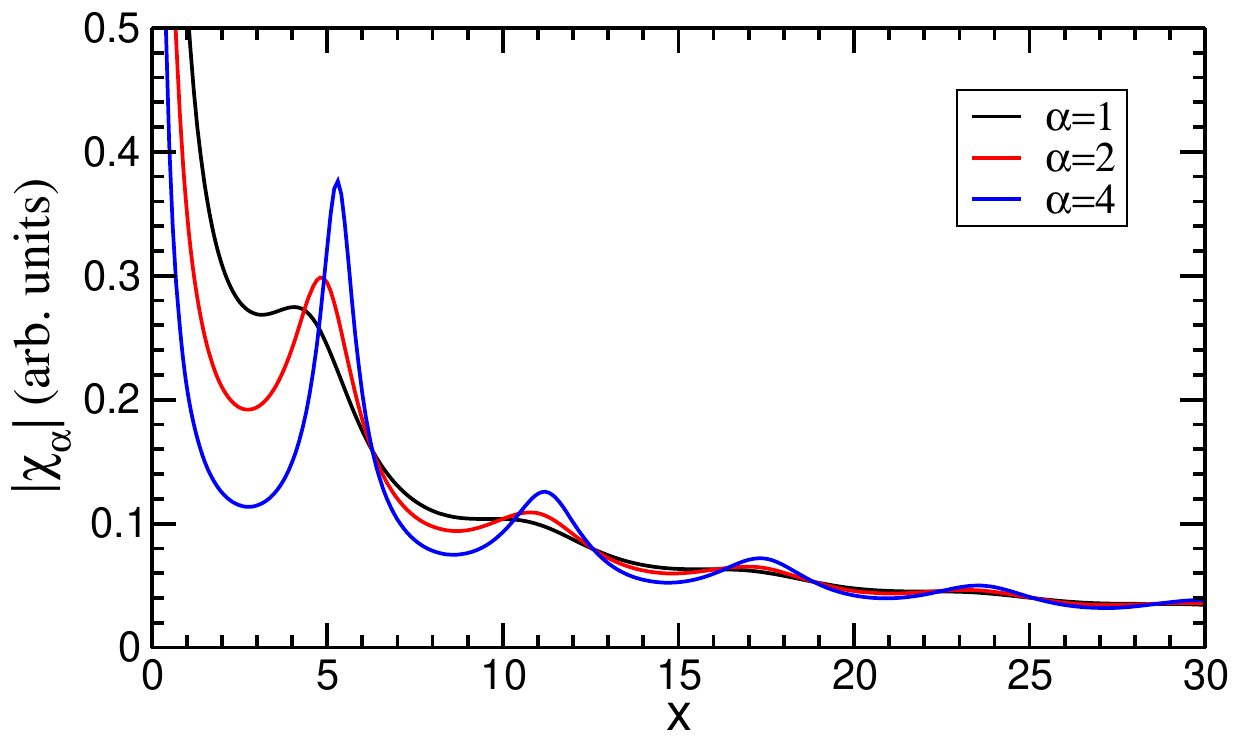}
    \caption{The absolute value of the susceptibility \eqref{eq:susz} is depicted. 
		The resonances are clearly visible as peaks, at least for not too small values
		of $\alpha$. The intrinsic width of these peaks results from the one-parameter form
		of the susceptibility. Strikingly, there occur many
		additional resonances after the first one which, however, become increasingly broader.}
    \label{fig:quant-b}
\end{figure}

To illustrate and corroborate this observation Fig.~\ref{fig:quant-b}b
displays the modulus of the susceptibility 
\be
\label{eq:susz}
\chi_\alpha(x) = \tau (-ix + \alpha(1-\exp(ix)))^{-1}
\ee
for three representative values of $\alpha$ given in the legend.
The striking feature is that the difference equation allows for several, presumably
infinitely many, resonances, all governed by the same time scale and parameter.
Figure \ref{fig:quant-b} clearly depicts this sequence of resonances with increasing
width for increasing $x$. The resonances occur approximately at equidistant frequencies.

\subsection{Signatures of multiple resonances}

The occurrence of many resonances as a series of essentially equidistant peaks
is a smoking gun signature of the response due to retarded relaxtion \eqref{eq:rlll}. 
The inertial LLG equation \eqref{eq:illg} allows only for two resonances because it
is second order in the derivatives yielding a second order characteristic polynomial
in frequency space. Thus, the detection of
many resonance peaks are a litmus text of the advocated model here.
Thus, it is highly intriguing that Unikandanunni et al.\ indeed found a sequence
of roughly equidistant resonance in three thin ferromagnetic Co films \cite{unika22}
of fcc, bcc and hcp lattices.

\begin{figure}
    \centering
    		\includegraphics[width=0.98\columnwidth,clip]{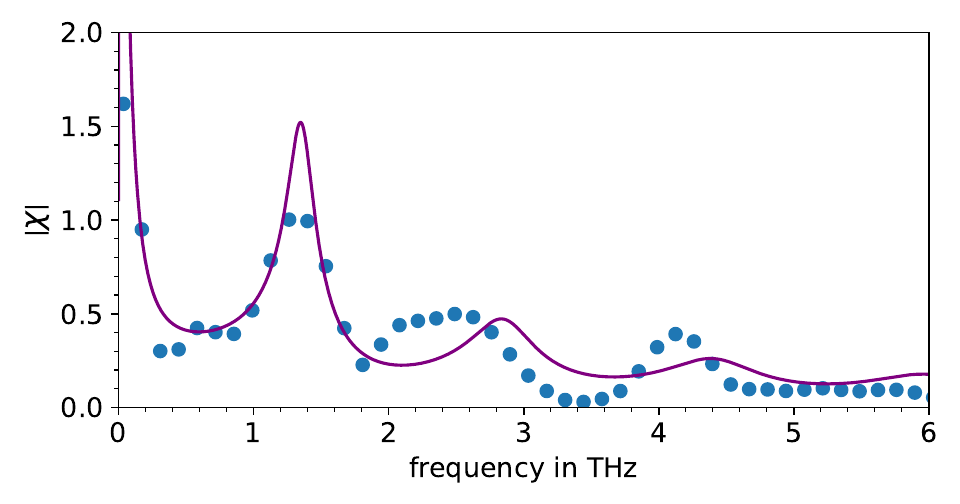}
    \caption{Fit to the measured susceptibility data from Unikandanunni et al.\ \cite{unika22}.
		The data was taken from a thin cobalt film of fcc structure. The theory is fitted
		to the first peak at finite frequency, the subsequent peaks are not fitted. In both data
		sets the response at low frequencies is deducted to highlight the high-frequency response.}
    \label{fig:mult-res}
\end{figure}

Figure \ref{fig:mult-res} displays a comparison of the data with a fit on the basis
of the retarded LLL equation \eqref{eq:rlll}. Since the experimental analysis 
deducted a slowly varying exponential decay we do so as well. Before computing the absolute
value of $\chi(\omega)$ we deduct the Fourier transform $A \tau_1/(1+i\omega\tau_1)$ 
where $\tau_1$ is taken from Ref.~\cite{unika22}. The value $\lambda$ is set to $0.015$ 
because the value in Ref.~\cite{unika22} is too large by an order of magnitude 
\cite{mukho23} although the precise value of the Gilbert damping is irrelevant for 
the ultrafast magnetization on the THz scale. The retardation time is fitted to be
$\tau=0.63$\,ps and $\tg=7.1$\,ps$^{-1}$ corresponding to $\alpha=4.5$.
Similar fits for the two other crystalline structures of Co thin films are provided in 
App.\ \ref{app:unika}.

The agreement is satisfying in view of the simplicity of the model. The key feature
is that the system displays multiple resonance peaks as they are a hallmark of
the retarded Lindblad-Landau-Lifshitz equation \eqref{eq:rlll}. This observation
corroborates the  LLL equation with and without retardation as promising
phenomenological extension of the LLG equation.

\section{Discussion}

In this article, we provide further evidence in favor of
a straightforward extension of the famous Landau-Lifshitz (LL) equation.
The topical motivation is the experimental observation of ultrafast modes in 
thin ferromagnetic films \cite{neera21,unika22,de23,de24}.
These modes were so far analyzed by Landau-Lifshitz-Gilbert (LLG) equations supplemented
by an inertial term, i.e., a second time derivative, leading to the inertial LLG (iLLG) equation. 

The here advocated extension of the LL equation
is based on the derivation of the LL equation from a microscopic Lindblad approach to a single spin.
A novel term occurs which induces dynamics at a higher energy scale; the
resulting equation is dubbed Lindblad-Landau-Lifshitz (LLL) equation \cite{uhrig25}.
The higher energy scale is linked to the relaxation rate $\gamma$ in the Lindblad dissipator
while the (Gilbert) damping parameter $\lambda$ in the LL or LLG equation is 
given by the ratio $\lambda=\hbar\gamma/J$.
Assuming instantaneous relaxation the additional term induces fast relaxation of the 
length of the magnetization. 

In order to explain the observed additional resonances, the iLLG equation is
often derived by means of a Taylor expansion in a retardation of the Gilbert damping.
Two challenges arise: (i) the next term in the expansion needs to be one to two orders
of magnitude larger than the leading one which speaks against truncating the Taylor 
expansion; (ii) the resonance peaks inherit their width from the Gilbert damping which is
far too small.

Taking up the idea of a retarded relaxation in this work, we analyze its effect
on the LLL equation leading to the retarded LLL equation \eqref{eq:rlll}.
We stress that no Taylor expansion of the retardation is applied. In the numerical and analytical
analysis of the dynamics described by the rLLL equation two key observations
are made: (a) the term at higher energy scale allows for oscillatory
modes in the THz range. The corresponding parameter fits well to the
energy scale set by the magnetic exchange field as it can be estimated from
the critical temperature. (b) These oscillatory modes come with a fairly
large decay rate of the same scale as the oscillation, i.e., the resonance
peaks are intrinsically broad. This fits very well to the experimental observation.

Furthermore, the analytic evaluation showed that a smoking gun feature of the
retarded LLL equations is the occurrence of several resonance peaks. Indeed,
experimental evidence in cobalt films indicates exactly this phenomenon.
Thereby, the advocated scenario is strongly corroborated.

While the quantitative agreement in the analysis of the leading, first resonance
is very good, see Fig. \ref{fig:comparison},
 the agreement of the subsequent resonance peaks is only qualitative, see
Fig.\ \ref{fig:mult-res}. In view of the simplicity of the advocated model
the agreement is satisfactory. For instance, a distribution of retardation times
suggests itself as a plausible extension. Still, the advocated model is phenomenologically
motivated and further, quantitative investigations of spin-bath models
are certainly called for, for instance in the spirit of Ref.\ \cite{hartm25}.

Yet, the introduced retarded LLL approach is a 
promising extension which can help to pave the way to 
atomistic spin simulations in the ultrafast regime.
Thereby, improved device simulation can come within reach
contributing to the development of magnonics and spintronics.

\backmatter

\bmhead{Supplementary information}

There is supplementary material available showing additional comparisons between
experimental and theoretical data in the appendices.

\bmhead{Acknowledgements}

We acknowledge helpful discussions with S. Bonetti and  J. Stolze; we 
thank V. Unikandanunni for the provision of data.

\section*{Declarations}

\begin{itemize}
\item Funding: Not applicable
\item Conflict of interest/Competing interests:
Both authors have no conflict of interest nor competing interests.
\item Ethics approval and consent to participate: Not applicable
\item Consent for publication:
Both authors consent to publish this article.
\item Data availability:
The data of the theoretical curves are available upon request. 
\item Materials availability:  Not applicable
\item Code availability: Not applicable
\item Author contribution: JB derived the general equations, produced
most of the figures and proof-read the manuscript. GSU had the main idea,
wrote the manuscript and provided the analytical understanding as well as two figures.
\end{itemize}

\begin{appendices}

\section{Analysis of further data by Neeraj et al.}\label{app:neera}

In the main text, the analysis of only one data set from Ref.\ \cite{neera21}
was presented which leaves the question whether the other data sets can
be described by the retarded Lindblad-Landau-Lifshitz (LLL) equations as well.

\begin{figure}
    \centering
    \includegraphics[width=0.98\columnwidth]{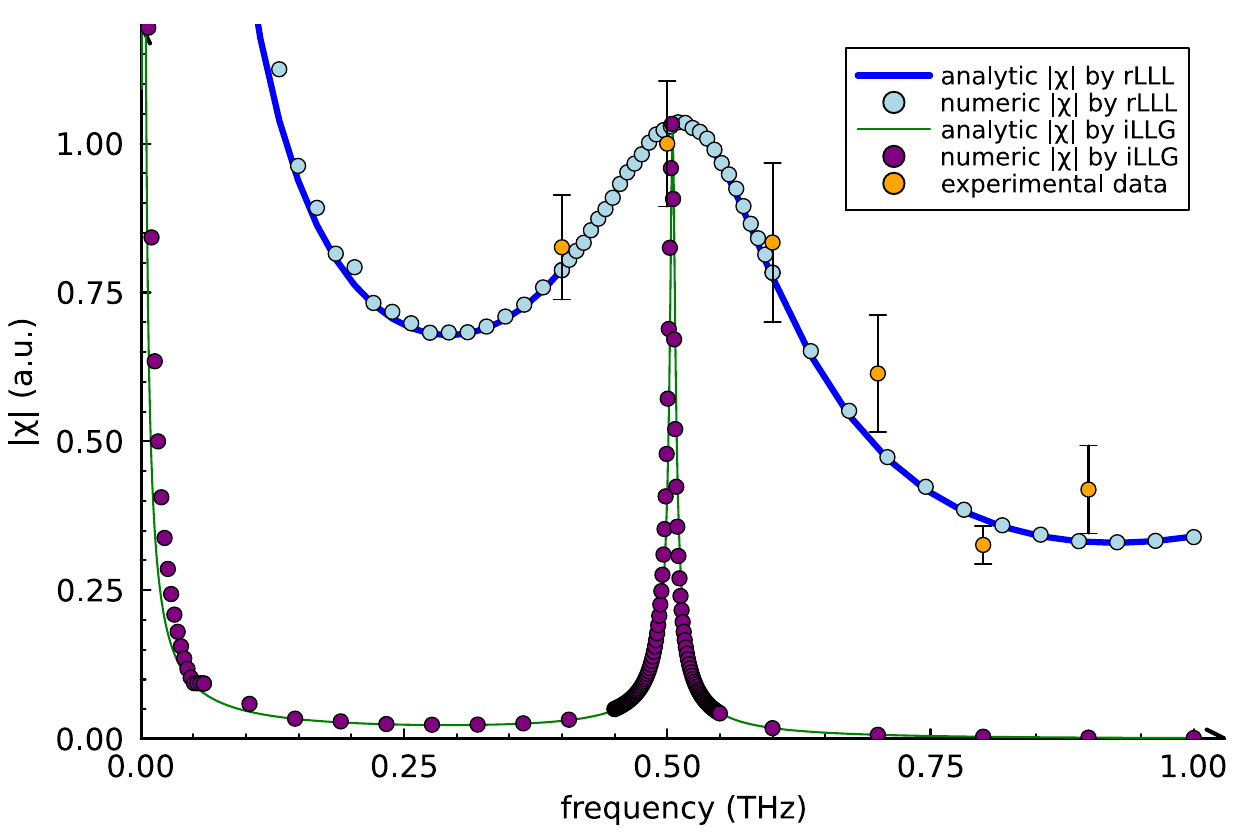}
    \caption{Dynamic susceptibility in arbitrary units 
		as measured in a CoFeB film in Ref.\ \cite{neera21} (yellow circles); as fitted with the 
		iLLG equation (3) with $\lambda=0.0044$ and $\ti=72$ps 
		(parameters from Ref.~\cite{neera21})
		from analytics (green line) and numerics (purple circles); 
		as fitted with retarded LLL equation \eqref{eq:rlll} with $\lambda=0.0044$, $\tau=1.50$\,ps, and 
		$\tilde\gamma = 1.30$\,ps$^{-1}$.}
    \label{fig:cofeb}
\end{figure}

Figure \ref{fig:cofeb} displays the data for CoFeB. All parameters are very similar
to the ones for epitaxial permalloy shown in the main text. An interesting comparison
results from an estimate of the relaxation rate $\tg=S\gamma$. As in the main text, we use
an estimate of $T_c\approx 1000$\,K \cite{jang20} and $S=0.72$  to obtain 
$\tg_\text{estimate}= $1.00\,ps$^{-1}$. This agrees with the fitted value only within 30\%, but still
illustrates that the parameters are physically absolutely reasonable.
Note as well  that we could argue that the bulk Curie temperature is the relevant one which 
increments the value for CoFeB to 1300-1400 K so that the deduced value of 
$\tg_\text{estimate}= $1.00\,ps$^{-1}$ would increase by about 30\% fitting again remarkably well
to the fitted values.

\begin{figure}
    \centering
    \includegraphics[width=0.98\columnwidth]{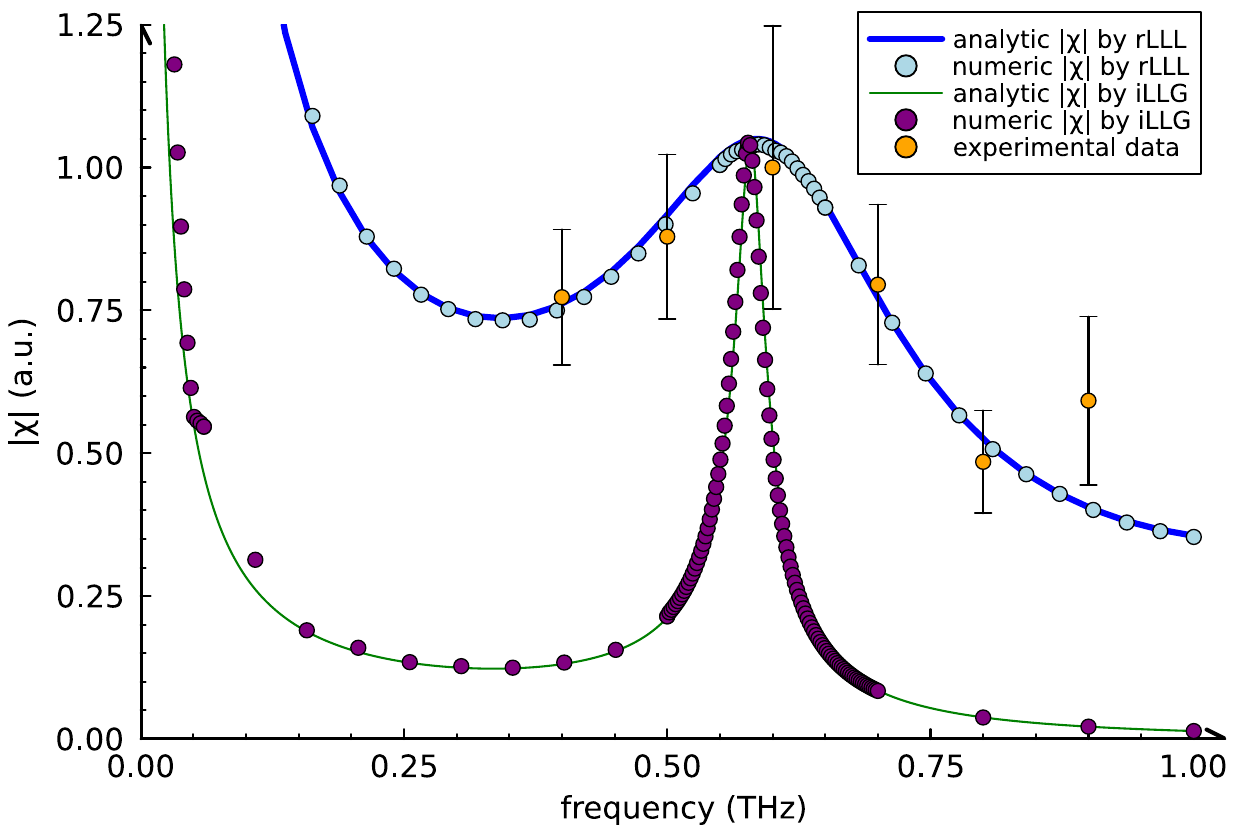}
    \caption{Dynamic susceptibility in arbitrary units 
		as measured in polycrystalline permalloy film in Ref.\ \cite{neera21} (yellow circles); a
		s fitted with the 
		iLLG equation (3) with $\lambda=0.023$ and $\ti=12$ps 
		(parameters from Ref.~\cite{neera21})
		from analytics (green line) and numerics (purple circles); 
		as fitted with retarded LLL equation \eqref{eq:rlll} with $\lambda=0.023$, $\tau=1.29$\,ps, and 
		$\tilde\gamma = 1.40$\,ps$^{-1}$.}
    \label{fig:polycrystal}
\end{figure}

Figure \ref{fig:polycrystal} displays the data for polycrystalline permalloy. 
All parameters are similar
to the ones for epitaxial permalloy shown in the main text. 
Note, however, the much larger value of the Gilbert damping which is about four times larger
than for the epitaxial permalloy. This makes the iLL fit broader by a factor of four, but
still much too narrow. Moreover, it is not clear why the damping is so much larger. 
A possible interpretation refers to the random orientation of the microcrystallites
which implies a substantial dephasing due to anisotropies. This dephasing
only mimics some intrinsic relaxation simulating a larger Gilbert damping.

The interesting comparison
results from an estimate of the relaxation rate $\tg=S\gamma$. As in the main text, we use
an estimate of $T_c\approx 875$\,K \cite{qader17} and $S=0.41$  to obtain 
$\tg_\text{estimate}= $5.61\,ps$^{-1}$. This agrees only within a factor of 4 with the 
fitted value. If, however,  we return to the above argument
that the Gilbert damping in the polycrystalline sample is not the intrinsic one, but
contains a strong spurious dephasing component,  then it is justified to use the 
Gilbert damping of the epitaxial sample of $\lambda=0.0058$ as the genuinely intrinsic
damping which implies $\tg_\text{estimate}= $1.41\,ps$^{-1}$. This value agrees again
astonishingly well with the fitted value.

We conclude that extremely good fits of the measured data are possible with perfectly
realistic parameter values for all three data sets from Ref.\ \cite{neera21}.

\section{Analysis of further data by Unikandanunni  et al.}\label{app:unika}

In the main text, the analysis of only one data set from Ref.\ \cite{unika22}
was presented.  The key observation was that in experiment as well
as in theory several resonance peaks occur at approximately equidistant frequencies
with decreasing amplitudes upon increasing frequency. Two more data sets
were measured and can be analyzed in the same way as in the main text.

\begin{figure}
    \centering
    \includegraphics[width=0.98\columnwidth]{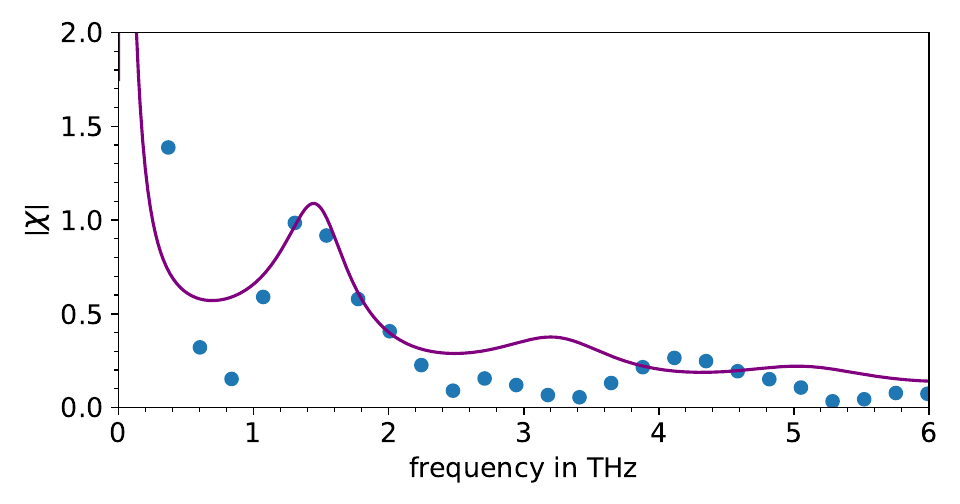}
    \caption{Measured data for a thin cobalt film with bcc structure compared to
		a fit based on the retarded Lindblad-Landau-Lifshitz equation \eqref{eq:rlll}. 
		The retardation is $\tau=0.55$\,ps while the Lindblad relaxation $\tg=4$\,ps$^{-1}$.}
    \label{fig:bcc}
\end{figure}

\begin{figure}
    \centering
    \includegraphics[width=0.98\columnwidth]{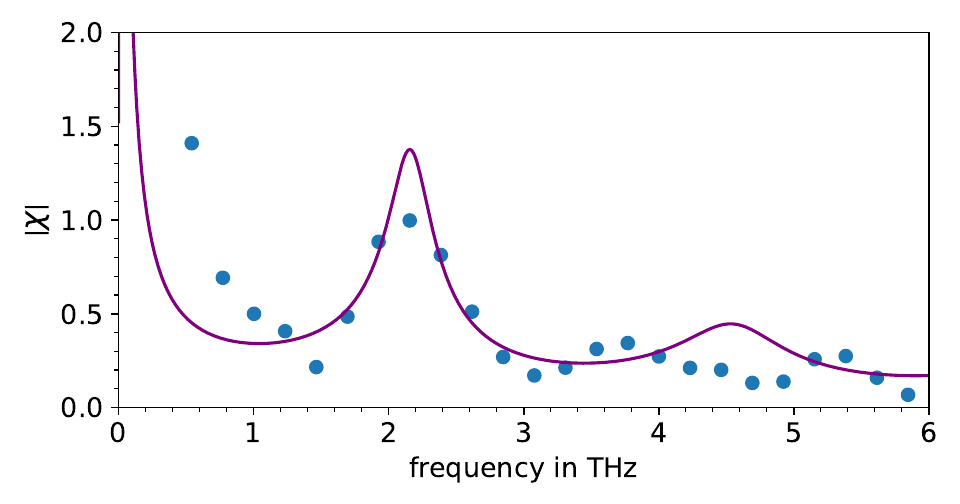}
    \caption{Measured data for a thin cobalt film with hcp structure compared to
		a fit based on the retarded Lindblad-Landau-Lifshitz equation \eqref{eq:rlll}. 
		The retardation is $\tau=0.395$\,ps while the Lindblad relaxation $\tg=11$\,ps$^{-1}$.}
    \label{fig:hcp}
\end{figure}

Figures \ref{fig:bcc} and \ref{fig:hcp} display the comparisons between experimental and
theoretical data. Note that in both sets an temporal exponential contribution (experiment)
or its Fourier transform (theory, see main text) have been deducted to emphasize the
ultrafast response. The theoretical data is fitted to the first resonance in the THz regime.
The agreement is not as satisfying as the one for the fcc film shown in the main text.
But the key feature to be kept in mind is clear here as well: 
Beyond first resonance further resonance peaks occur, both in experiment and in theory.
Improved theoretical and experimental analysis is called for.

\section{Estimate of spin values}

In the itinerant ferromagnets considered no 
quantized spin value $S$ can be determined. But it is possible to estimate
a physically plausible value. For polycrystalline permalloy Ni$_{0.81}$Fe$_{0.19}$, the argument
runs as follows.

The average atomic mass $m=58.1$\,g/mole results from the 
weighted atomic masses Ni with $58.7$\,g/mole and Fe with $55.8$\,g/mole.
Combined with the mass density of permalloy of $\rho=8.91$\,g/cm$^3$ we obtain
a volume per mole of $v=m/\rho=6.52\cdot10^{-8}$\,m$^3$. Then the dipole moment
of a single average atom $d$  is deduced from the saturation magnetization $M_S$
by $d=M_S v/N_A=7.99\cdot 10^{-24}$\,Am$^2$ where $N_A$ is Avogadro's constant.
Comparing this to $d=g\mu_\text{B} S$ we deduce $S\approx 0.41$.

For epitaxial permalloy, we analogously obtain $S\approx 0.46$ and for CoFeB
we obtain $S\approx 0.72$. These values are used in the main text to link
the critical temperature to the summed exchange coupling $J$ and thereby
eventually $\tg$.

\end{appendices}



\begin{thebibliography}{43}
\ifx \bisbn   \undefined \def \bisbn  #1{ISBN #1}\fi
\ifx \binits  \undefined \def \binits#1{#1}\fi
\ifx \bauthor  \undefined \def \bauthor#1{#1}\fi
\ifx \batitle  \undefined \def \batitle#1{#1}\fi
\ifx \bjtitle  \undefined \def \bjtitle#1{#1}\fi
\ifx \bvolume  \undefined \def \bvolume#1{\textbf{#1}}\fi
\ifx \byear  \undefined \def \byear#1{#1}\fi
\ifx \bissue  \undefined \def \bissue#1{#1}\fi
\ifx \bfpage  \undefined \def \bfpage#1{#1}\fi
\ifx \blpage  \undefined \def \blpage #1{#1}\fi
\ifx \burl  \undefined \def \burl#1{\textsf{#1}}\fi
\ifx \doiurl  \undefined \def \doiurl#1{\url{https://doi.org/#1}}\fi
\ifx \betal  \undefined \def \betal{\textit{et al.}}\fi
\ifx \binstitute  \undefined \def \binstitute#1{#1}\fi
\ifx \binstitutionaled  \undefined \def \binstitutionaled#1{#1}\fi
\ifx \bctitle  \undefined \def \bctitle#1{#1}\fi
\ifx \beditor  \undefined \def \beditor#1{#1}\fi
\ifx \bpublisher  \undefined \def \bpublisher#1{#1}\fi
\ifx \bbtitle  \undefined \def \bbtitle#1{#1}\fi
\ifx \bedition  \undefined \def \bedition#1{#1}\fi
\ifx \bseriesno  \undefined \def \bseriesno#1{#1}\fi
\ifx \blocation  \undefined \def \blocation#1{#1}\fi
\ifx \bsertitle  \undefined \def \bsertitle#1{#1}\fi
\ifx \bsnm \undefined \def \bsnm#1{#1}\fi
\ifx \bsuffix \undefined \def \bsuffix#1{#1}\fi
\ifx \bparticle \undefined \def \bparticle#1{#1}\fi
\ifx \barticle \undefined \def \barticle#1{#1}\fi
\bibcommenthead
\ifx \bconfdate \undefined \def \bconfdate #1{#1}\fi
\ifx \botherref \undefined \def \botherref #1{#1}\fi
\ifx \url \undefined \def \url#1{\textsf{#1}}\fi
\ifx \bchapter \undefined \def \bchapter#1{#1}\fi
\ifx \bbook \undefined \def \bbook#1{#1}\fi
\ifx \bcomment \undefined \def \bcomment#1{#1}\fi
\ifx \oauthor \undefined \def \oauthor#1{#1}\fi
\ifx \citeauthoryear \undefined \def \citeauthoryear#1{#1}\fi
\ifx \endbibitem  \undefined \def \endbibitem {}\fi
\ifx \bconflocation  \undefined \def \bconflocation#1{#1}\fi
\ifx \arxivurl  \undefined \def \arxivurl#1{\textsf{#1}}\fi
\csname PreBibitemsHook\endcsname

\bibitem[\protect\citeauthoryear{Chappert et~al.}{2007}]{chapp07}
\begin{barticle}
\bauthor{\bsnm{Chappert}, \binits{C.}},
\bauthor{\bsnm{Fert}, \binits{A.}},
\bauthor{\bsnm{Van~Dau}, \binits{F.}}:
\batitle{The emergence of spin electronics in data storage}.
\bjtitle{Nature Materials}
\bvolume{6},
\bfpage{813}
(\byear{2007})
\doiurl{10.1038/nmat2024}
\end{barticle}
\endbibitem

\bibitem[\protect\citeauthoryear{Gomonay and Loktev}{2014}]{gomon14}
\begin{barticle}
\bauthor{\bsnm{Gomonay}, \binits{E.V.}},
\bauthor{\bsnm{Loktev}, \binits{V.M.}}:
\batitle{Spintronics of antiferromagnetic systems (review article)}.
\bjtitle{Low Temperature Physics}
\bvolume{40},
\bfpage{17}
(\byear{2014})
\doiurl{10.1063/1.4862467}
\end{barticle}
\endbibitem

\bibitem[\protect\citeauthoryear{Chumak et~al.}{2015}]{chuma15}
\begin{barticle}
\bauthor{\bsnm{Chumak}, \binits{A.V.}},
\bauthor{\bsnm{Vasyuchka}, \binits{V.I.}},
\bauthor{\bsnm{Serga}, \binits{A.A.}},
\bauthor{\bsnm{Hillebrands}, \binits{B.}}:
\batitle{Magnon spintronics}.
\bjtitle{Nature Physics}
\bvolume{11},
\bfpage{453}
(\byear{2015})
\doiurl{10.1038/NPHYS3347}
\end{barticle}
\endbibitem

\bibitem[\protect\citeauthoryear{Barman et~al.}{2021}]{barma21}
\begin{barticle}
\bauthor{\bsnm{Barman}, \binits{A.}},
\bauthor{\bsnm{Gubbiotti}, \binits{G.}},
\bauthor{\bsnm{Ladak}, \binits{S.}},
\bauthor{\bsnm{Adeyeye}, \binits{A.O.}},
\bauthor{\bsnm{Krawczyk}, \binits{M.}},
\bauthor{\bsnm{Gr\"afe}, \binits{J.}},
\bauthor{\bsnm{Adelmann}, \binits{C.}},
\bauthor{\bsnm{Cotofana}, \binits{S.}},
\bauthor{\bsnm{Naeemi}, \binits{A.}},
\bauthor{\bsnm{Vasyuchka}, \binits{V.I.}},
\bauthor{\bsnm{Hillebrands}, \binits{B.}},
\bauthor{\bsnm{Nikitov}, \binits{S.A.}},
\bauthor{\bsnm{Yu}, \binits{H.}},
\bauthor{\bsnm{Grundler}, \binits{D.}},
\bauthor{\bsnm{Sadovnikov}, \binits{A.}},
\bauthor{\bsnm{Grachev}, \binits{A.A.}},
\bauthor{\bsnm{Sheshukova}, \binits{S.E.}},
\bauthor{\bsnm{Duquesne}, \binits{J.-Y.}},
\bauthor{\bsnm{Marangolo}, \binits{M.}},
\bauthor{\bsnm{Gyorgy}, \binits{C.}},
\bauthor{\bsnm{Porod}, \binits{W.}},
\bauthor{\bsnm{Demidov}, \binits{V.E.}},
\bauthor{\bsnm{Urazhdin}, \binits{S.}},
\bauthor{\bsnm{Demokritov}, \binits{S.}},
\bauthor{\bsnm{Albisetti}, \binits{E.}},
\bauthor{\bsnm{Petti}, \binits{D.}},
\bauthor{\bsnm{Bertacco}, \binits{R.}},
\bauthor{\bsnm{Schulteiss}, \binits{H.}},
\bauthor{\bsnm{Kruglyak}, \binits{V.V.}},
\bauthor{\bsnm{Poimanov}, \binits{V.D.}},
\bauthor{\bsnm{Sahoo}, \binits{A.K.}},
\bauthor{\bsnm{Sinha}, \binits{J.}},
\bauthor{\bsnm{Yang}, \binits{H.}},
\bauthor{\bsnm{Muenzenberg}, \binits{M.}},
\bauthor{\bsnm{Moriyama}, \binits{T.}},
\bauthor{\bsnm{Mizukami}, \binits{S.}},
\bauthor{\bsnm{Landeros}, \binits{P.}},
\bauthor{\bsnm{Gallardo}, \binits{R.A.}},
\bauthor{\bsnm{Carlotti}, \binits{G.}},
\bauthor{\bsnm{Kim}, \binits{J.-V.}},
\bauthor{\bsnm{Stamps}, \binits{R.L.}},
\bauthor{\bsnm{Camley}, \binits{R.E.}},
\bauthor{\bsnm{Rana}, \binits{B.}},
\bauthor{\bsnm{Otani}, \binits{Y.}},
\bauthor{\bsnm{Yu}, \binits{W.}},
\bauthor{\bsnm{Yu}, \binits{T.}},
\bauthor{\bsnm{Bauer}, \binits{G.E.W.}},
\bauthor{\bsnm{Back}, \binits{C.H.}},
\bauthor{\bsnm{Uhrig}, \binits{G.S.}},
\bauthor{\bsnm{Dobrovolskiy}, \binits{O.V.}},
\bauthor{\bsnm{Dijken}, \binits{S.}},
\bauthor{\bsnm{Budinska}, \binits{B.}},
\bauthor{\bsnm{Qin}, \binits{H.}},
\bauthor{\bsnm{Chumak}, \binits{A.}},
\bauthor{\bsnm{Khitun}, \binits{A.}},
\bauthor{\bsnm{Nikonov}, \binits{D.E.}},
\bauthor{\bsnm{Young}, \binits{I.A.}},
\bauthor{\bsnm{Zingsem}, \binits{B.}},
\bauthor{\bsnm{Winklhofer}, \binits{M.}}:
\batitle{The 2021 {Magnonics Roadmap}}.
\bjtitle{J. Phys. Cond. Mat.}
\bvolume{33},
\bfpage{413001}
(\byear{2021})
\doiurl{10.1088/1361-648X/abec1a}
\end{barticle}
\endbibitem

\bibitem[\protect\citeauthoryear{Flebus et~al.}{2024}]{flebu24}
\begin{barticle}
\bauthor{\bsnm{Flebus}, \binits{B.}},
\bauthor{\bsnm{Grundler}, \binits{D.}},
\bauthor{\bsnm{Rana}, \binits{B.}},
\bauthor{\bsnm{Otani}, \binits{Y.}},
\bauthor{\bsnm{Barsukov}, \binits{I.}},
\bauthor{\bsnm{Barman}, \binits{A.}},
\bauthor{\bsnm{Gubbiotti}, \binits{G.}},
\bauthor{\bsnm{Landeros}, \binits{P.}},
\bauthor{\bsnm{Akerman}, \binits{J.}},
\bauthor{\bsnm{Ebels}, \binits{U.}},
\bauthor{\bsnm{Pirro}, \binits{P.}},
\bauthor{\bsnm{Demidov}, \binits{V.E.}},
\bauthor{\bsnm{Schultheiss}, \binits{K.}},
\bauthor{\bsnm{Csaba}, \binits{G.}},
\bauthor{\bsnm{Wang}, \binits{Q.}},
\bauthor{\bsnm{Ciubotaru}, \binits{F.}},
\bauthor{\bsnm{Nikonov}, \binits{D.E.}},
\bauthor{\bsnm{Che}, \binits{P.}},
\bauthor{\bsnm{Hertel}, \binits{R.}},
\bauthor{\bsnm{Ono}, \binits{T.}},
\bauthor{\bsnm{Afanasiev}, \binits{D.}},
\bauthor{\bsnm{Mentink}, \binits{J.}},
\bauthor{\bsnm{Rasing}, \binits{T.}},
\bauthor{\bsnm{Hillebrands}, \binits{B.}},
\bauthor{\bsnm{Kusminskiy}, \binits{S.V.}},
\bauthor{\bsnm{Zhang}, \binits{W.}},
\bauthor{\bsnm{Du}, \binits{C.R.}},
\bauthor{\bsnm{Finco}, \binits{A.}},
\bauthor{\bsnm{Sar}, \binits{T.}},
\bauthor{\bsnm{Luo}, \binits{Y.K.}},
\bauthor{\bsnm{Shiota}, \binits{Y.}},
\bauthor{\bsnm{Sklenar}, \binits{J.}},
\bauthor{\bsnm{Yu}, \binits{T.}},
\bauthor{\bsnm{Rao}, \binits{J.}}:
\batitle{The 2024 magnonics roadmap}.
\bjtitle{Journal of Physics: Condensed Matter}
\bvolume{36},
\bfpage{363501}
(\byear{2024})
\doiurl{10.1088/1361-648X/ad399c}
\end{barticle}
\endbibitem

\bibitem[\protect\citeauthoryear{Hartmann et~al.}{2025}]{hartm25}
\begin{botherref}
\oauthor{\bsnm{Hartmann}, \binits{F.}},
\oauthor{\bsnm{Unikandanunni}, \binits{V.}},
\oauthor{\bsnm{Bargheer}, \binits{M.}},
\oauthor{\bsnm{Fullerton}, \binits{E.E.}},
\oauthor{\bsnm{Bonetti}, \binits{S.}},
\oauthor{\bsnm{Anders}, \binits{J.}}:
Intrinsic non-markovian magnetisation dynamics.
arXiv,
2512--07378
(2025)
\doiurl{10.48550/arXiv.2512.07378}
\end{botherref}
\endbibitem

\bibitem[\protect\citeauthoryear{Lenzing et~al.}{2025}]{lenzi25}
\begin{barticle}
\bauthor{\bsnm{Lenzing}, \binits{N.}},
\bauthor{\bsnm{Kr\"uger}, \binits{D.}},
\bauthor{\bsnm{Potthoff}, \binits{M.}}:
\batitle{Microscopic theory of spin friction and dissipative spin dynamics}.
\bjtitle{Physical Review B}
\bvolume{111},
\bfpage{014402}
(\byear{2025})
\doiurl{10.1103/PhysRevB.111.014402}
\end{barticle}
\endbibitem

\bibitem[\protect\citeauthoryear{Landau and Lifshitz}{1935}]{landa35}
\begin{barticle}
\bauthor{\bsnm{Landau}, \binits{L.D.}},
\bauthor{\bsnm{Lifshitz}, \binits{E.M.}}:
\batitle{On the theory of the dispersion of magnetic permeability in
  ferromagnetic bodies}.
\bjtitle{Physikalische Zeitschrift der Sowjetunion}
\bvolume{8},
\bfpage{153}
(\byear{1935})
\doiurl{10.1016/b978-0-08-010586-4.50023-7}
\end{barticle}
\endbibitem

\bibitem[\protect\citeauthoryear{Gilbert}{2004}]{gilbe04}
\begin{barticle}
\bauthor{\bsnm{Gilbert}, \binits{T.L.}}:
\batitle{A {Phenomenological Theory of Damping in Ferromagnetic Materials}}.
\bjtitle{IEEE Transactions on Magnetics}
\bvolume{40},
\bfpage{3443}
(\byear{2004})
\doiurl{10.1109/TMAG.2004.836740}
\end{barticle}
\endbibitem

\bibitem[\protect\citeauthoryear{Beaurepaire et~al.}{1996}]{beaur96}
\begin{barticle}
\bauthor{\bsnm{Beaurepaire}, \binits{E.}},
\bauthor{\bsnm{Merle}, \binits{J.-C.}},
\bauthor{\bsnm{Daunois}, \binits{A.}},
\bauthor{\bsnm{Bigot}, \binits{J.-Y.}}:
\batitle{Ultrafast spin dynamics in ferromagnetic nickel}.
\bjtitle{Physical Review Letters}
\bvolume{76},
\bfpage{4250}
(\byear{1996})
\doiurl{10.1103/PhysRevLett.76.4250}
\end{barticle}
\endbibitem

\bibitem[\protect\citeauthoryear{Ciornei et~al.}{2011}]{ciorn11}
\begin{barticle}
\bauthor{\bsnm{Ciornei}, \binits{M.-C.}},
\bauthor{\bsnm{Rub\'i}, \binits{J.M.}},
\bauthor{\bsnm{Wegrowe}, \binits{J.-E.}}:
\batitle{Magnetization dynamics in the inertial regime: Nutation predicted at
  short time scales}.
\bjtitle{Phys. Rev. B}
\bvolume{83},
\bfpage{020410}
(\byear{2011})
\doiurl{10.1103/PhysRevB.83.020410}
\end{barticle}
\endbibitem

\bibitem[\protect\citeauthoryear{Wegrowe and Ciornei}{2012}]{wegro12}
\begin{barticle}
\bauthor{\bsnm{Wegrowe}, \binits{J.-E.}},
\bauthor{\bsnm{Ciornei}, \binits{M.-C.}}:
\batitle{Magnetization dynamics, gyromagnetic relation, and inertial effects}.
\bjtitle{Am. J. Phys.}
\bvolume{80},
\bfpage{607}
(\byear{2012})
\doiurl{10.1119/1.4709188}
\end{barticle}
\endbibitem

\bibitem[\protect\citeauthoryear{Olive et~al.}{2012a}]{olive12}
\begin{barticle}
\bauthor{\bsnm{Olive}, \binits{E.}},
\bauthor{\bsnm{Lansac}, \binits{Y.}},
\bauthor{\bsnm{Wegrowe}, \binits{J.-E.}}:
\batitle{Beyond ferromagnetic resonance: The inertial regime of the
  magnetization}.
\bjtitle{Appl. Phys. Lett.}
\bvolume{100},
\bfpage{192407}
(\byear{2012})
\doiurl{10.1063/1.4712056}
\end{barticle}
\endbibitem

\bibitem[\protect\citeauthoryear{Olive et~al.}{2012b}]{olive15}
\begin{barticle}
\bauthor{\bsnm{Olive}, \binits{E.}},
\bauthor{\bsnm{Lansac}, \binits{Y.}},
\bauthor{\bsnm{Meyer}, \binits{M.}},
\bauthor{\bsnm{Hayoun}, \binits{M.}},
\bauthor{\bsnm{Wegrowe}, \binits{J.-E.}}:
\batitle{Deviation from the landau-lifshitz-gilbert equation in the inertial
  regime of the magnetization}.
\bjtitle{J. Appl. Phys.}
\bvolume{117},
\bfpage{213904}
(\byear{2012})
\doiurl{10.1063/1.4921908}
\end{barticle}
\endbibitem

\bibitem[\protect\citeauthoryear{Bhattacharjee et~al.}{2012}]{bhatt12}
\begin{barticle}
\bauthor{\bsnm{Bhattacharjee}, \binits{S.}},
\bauthor{\bsnm{Nordstr\"om}, \binits{L.}},
\bauthor{\bsnm{Fransson}, \binits{J.}}:
\batitle{Atomistic {Spin Dynamic Method with both Damping and Moment of Inertia
  Effects Included from First Principles}}.
\bjtitle{Physical Review Letters}
\bvolume{108},
\bfpage{057204}
(\byear{2012})
\doiurl{10.1103/PhysRevLett.108.057204}
\end{barticle}
\endbibitem

\bibitem[\protect\citeauthoryear{Kikuchi and Tatara}{2015}]{kikuc15}
\begin{barticle}
\bauthor{\bsnm{Kikuchi}, \binits{T.}},
\bauthor{\bsnm{Tatara}, \binits{G.}}:
\batitle{Spin dynamics with inertia in metallic ferromagnets}.
\bjtitle{Physical Review B}
\bvolume{92},
\bfpage{184410}
(\byear{2015})
\doiurl{10.1103/PhysRevB.92.184410}
\end{barticle}
\endbibitem

\bibitem[\protect\citeauthoryear{Sayad et~al.}{2016}]{sayad16b}
\begin{barticle}
\bauthor{\bsnm{Sayad}, \binits{M.}},
\bauthor{\bsnm{Rausch}, \binits{R.}},
\bauthor{\bsnm{Potthoff}, \binits{M.}}:
\batitle{Inertia effects in the real-time dynamics of a quantum spin coupled to
  a fermi sea}.
\bjtitle{Europhysics Letters}
\bvolume{116},
\bfpage{17001}
(\byear{2016})
\doiurl{10.1209/0295-5075/116/17001}
\end{barticle}
\endbibitem

\bibitem[\protect\citeauthoryear{F\"ahnle et~al.}{2011}]{fahnl11}
\begin{barticle}
\bauthor{\bsnm{F\"ahnle}, \binits{M.}},
\bauthor{\bsnm{Steiauf}, \binits{D.}},
\bauthor{\bsnm{Illg}, \binits{C.}}:
\batitle{Generalized {Gilbert equation including inertial damping: Derivation
  from an extended breathing Fermi} surface model}.
\bjtitle{Physical Review B}
\bvolume{84},
\bfpage{172403}
(\byear{2011})
\doiurl{10.1103/PhysRevB.84.172403}
\end{barticle}
\endbibitem

\bibitem[\protect\citeauthoryear{F\"ahnle et~al.}{2013}]{fahnl11err}
\begin{barticle}
\bauthor{\bsnm{F\"ahnle}, \binits{M.}},
\bauthor{\bsnm{Steiauf}, \binits{D.}},
\bauthor{\bsnm{Illg}, \binits{C.}}:
\batitle{Erratum: {Generalized Gilbert equation including inertial damping:
  Derivation from an extended breathing Fermi surface model [Phys. Rev. B 84,
  172403 (2011)]}}.
\bjtitle{Physical Review B}
\bvolume{88},
\bfpage{219905}
(\byear{2013})
\doiurl{10.1103/PhysRevB.88.219905}
\end{barticle}
\endbibitem

\bibitem[\protect\citeauthoryear{Bajpai and Nikoli\'c}{2019}]{bajpa19}
\begin{barticle}
\bauthor{\bsnm{Bajpai}, \binits{U.}},
\bauthor{\bsnm{Nikoli\'c}, \binits{B.K.}}:
\batitle{Time-retarded damping and magnetic inertia in the
  {Landau-Lifshitz-Gilbert equation self-consistently coupled to electronic
  time-dependent nonequilibrium Green} functions}.
\bjtitle{Physical Review B}
\bvolume{99},
\bfpage{134409}
(\byear{2019})
\doiurl{10.1103/PhysRevB.99.134409}
\end{barticle}
\endbibitem

\bibitem[\protect\citeauthoryear{Thonig et~al.}{2017}]{thoni17}
\begin{barticle}
\bauthor{\bsnm{Thonig}, \binits{D.}},
\bauthor{\bsnm{Eriksson}, \binits{O.}},
\bauthor{\bsnm{Pereiro}, \binits{M.}}:
\batitle{Magnetic moment of inertia within the torque-torque correlation
  model}.
\bjtitle{Sci. Rep.}
\bvolume{7},
\bfpage{931}
(\byear{2017})
\doiurl{10.1038/s41598-017-01081-z}
\end{barticle}
\endbibitem

\bibitem[\protect\citeauthoryear{Mondal et~al.}{2017}]{monda17}
\begin{barticle}
\bauthor{\bsnm{Mondal}, \binits{R.}},
\bauthor{\bsnm{Berritta}, \binits{M.}},
\bauthor{\bsnm{Nandy}, \binits{A.K.}},
\bauthor{\bsnm{Oppeneer}, \binits{P.M.}}:
\batitle{Relativistic theory of magnetic inertia in ultrafast spin dynamics}.
\bjtitle{Physical Review B}
\bvolume{96},
\bfpage{024425}
(\byear{2017})
\doiurl{10.1103/PhysRevB.96.024425}
\end{barticle}
\endbibitem

\bibitem[\protect\citeauthoryear{Mondal et~al.}{2018}]{monda18}
\begin{barticle}
\bauthor{\bsnm{Mondal}, \binits{R.}},
\bauthor{\bsnm{Berritta}, \binits{M.}},
\bauthor{\bsnm{Oppeneer}, \binits{P.M.}}:
\batitle{Generalisation of gilbert damping and magnetic inertia parameter as a
  series of higher-order relativistic terms}.
\bjtitle{Journal of Physics: Condensed Matter}
\bvolume{30},
\bfpage{265801}
(\byear{2018})
\doiurl{10.1088/1361-648X/aac5a2}
\end{barticle}
\endbibitem

\bibitem[\protect\citeauthoryear{Mondal et~al.}{2023}]{monda23b}
\begin{barticle}
\bauthor{\bsnm{Mondal}, \binits{R.}},
\bauthor{\bsnm{R\'ozsa}, \binits{L.}},
\bauthor{\bsnm{Farle}, \binits{M.}},
\bauthor{\bsnm{Oppeneer}, \binits{P.M.}},
\bauthor{\bsnm{Nowak}, \binits{U.}},
\bauthor{\bsnm{Cherkasskii}, \binits{M.}}:
\batitle{Inertial effects in ultrafast spin dynamics}.
\bjtitle{J. Mag. Mag. Mat.}
\bvolume{579},
\bfpage{170830}
(\byear{2023})
\doiurl{10.1016/j.jmmm.2023.170830}
\end{barticle}
\endbibitem

\bibitem[\protect\citeauthoryear{Quarenta et~al.}{2024}]{quare24}
\begin{barticle}
\bauthor{\bsnm{Quarenta}, \binits{M.G.}},
\bauthor{\bsnm{Tharmalingam}, \binits{M.}},
\bauthor{\bsnm{Ludwig}, \binits{T.}},
\bauthor{\bsnm{Yuan}, \binits{H.Y.}},
\bauthor{\bsnm{Karwacki}, \binits{L.}},
\bauthor{\bsnm{Verstraten}, \binits{R.C.}},
\bauthor{\bsnm{Duine}, \binits{R.A.}}:
\batitle{Bath-induced spin inertia}.
\bjtitle{Physical Review Letters}
\bvolume{133},
\bfpage{136701}
(\byear{2024})
\doiurl{10.1103/PhysRevLett.133.136701}
\end{barticle}
\endbibitem

\bibitem[\protect\citeauthoryear{Bastardis et~al.}{2018}]{basta18}
\begin{barticle}
\bauthor{\bsnm{Bastardis}, \binits{R.}},
\bauthor{\bsnm{Vernay}, \binits{F.}},
\bauthor{\bsnm{Kachkachi}, \binits{H.}}:
\batitle{Magnetization nutation induced by surface effects in nanomagnets}.
\bjtitle{Physical Review B}
\bvolume{98},
\bfpage{165444}
(\byear{2018})
\doiurl{10.1103/PhysRevB.98.165444}
\end{barticle}
\endbibitem

\bibitem[\protect\citeauthoryear{Reyes-Osorio and Nikoli\'c}{2025}]{reyes25}
\begin{barticle}
\bauthor{\bsnm{Reyes-Osorio}, \binits{F.}},
\bauthor{\bsnm{Nikoli\'c}, \binits{B.K.}}:
\batitle{Optically {Induced Magnetic Inertia and Magnons from Non-Markovian
  Extension of the Landau-Lifshitz-Gilbert Equation}}.
\bjtitle{Physical Review Letters}
\bvolume{135},
\bfpage{246701}
(\byear{2025})
\doiurl{10.1103/sl4k-pcvq}
\end{barticle}
\endbibitem

\bibitem[\protect\citeauthoryear{Kimel et~al.}{2009}]{kimel09}
\begin{barticle}
\bauthor{\bsnm{Kimel}, \binits{A.V.}},
\bauthor{\bsnm{Ivanov}, \binits{B.A.}},
\bauthor{\bsnm{Pisarev}, \binits{R.V.}},
\bauthor{\bsnm{Usachev}, \binits{P.A.}},
\bauthor{\bsnm{Kirilyuk}, \binits{A.}},
\bauthor{\bsnm{Rasing}, \binits{T.}}:
\batitle{Inertia-driven spin switching in antiferromagnets}.
\bjtitle{Nature Physics}
\bvolume{5},
\bfpage{727}
(\byear{2009})
\doiurl{10.1038/nphys1369}
\end{barticle}
\endbibitem

\bibitem[\protect\citeauthoryear{Heisterkamp et~al.}{2015}]{heist15}
\begin{barticle}
\bauthor{\bsnm{Heisterkamp}, \binits{F.}},
\bauthor{\bsnm{Zhukov}, \binits{E.A.}},
\bauthor{\bsnm{Greilich}, \binits{A.}},
\bauthor{\bsnm{Yakovlev}, \binits{D.R.}},
\bauthor{\bsnm{Korenev}, \binits{V.L.}},
\bauthor{\bsnm{Pawlis}, \binits{A.}},
\bauthor{\bsnm{Bayer}, \binits{M.}}:
\batitle{Longitudinal and transverse spin dynamics of donor-bound electrons in
  fluorine-doped {ZnSe}: Spin inertia versus {Hanle} effect}.
\bjtitle{Physical Review B}
\bvolume{91},
\bfpage{235432}
(\byear{2015})
\doiurl{10.1103/PhysRevB.91.235432}
\end{barticle}
\endbibitem

\bibitem[\protect\citeauthoryear{Schering et~al.}{2019}]{scher19}
\begin{barticle}
\bauthor{\bsnm{Schering}, \binits{P.}},
\bauthor{\bsnm{Uhrig}, \binits{G.S.}},
\bauthor{\bsnm{Smirnov}, \binits{D.S.}}:
\batitle{Spin inertia and polarization recovery in quantum dots: Role of
  pumping strength and resonant spin amplification}.
\bjtitle{Physical Review Research}
\bvolume{1},
\bfpage{033189}
(\byear{2019})
\doiurl{10.1103/PhysRevResearch.1.033189}
\end{barticle}
\endbibitem

\bibitem[\protect\citeauthoryear{Li et~al.}{2015}]{li15d}
\begin{barticle}
\bauthor{\bsnm{Li}, \binits{Y.}},
\bauthor{\bsnm{Barra}, \binits{A.-L.}},
\bauthor{\bsnm{Auffret}, \binits{S.}},
\bauthor{\bsnm{Ebels}, \binits{U.}},
\bauthor{\bsnm{Bailey}, \binits{W.E.}}:
\batitle{Inertial terms to magnetization dynamics in ferromagnetic thin films}.
\bjtitle{Physical Review B}
\bvolume{92},
\bfpage{140413}
(\byear{2015})
\doiurl{10.1103/PhysRevB.92.140413}
\end{barticle}
\endbibitem

\bibitem[\protect\citeauthoryear{Neeraj et~al.}{2021}]{neera21}
\begin{barticle}
\bauthor{\bsnm{Neeraj}, \binits{K.}},
\bauthor{\bsnm{Awari}, \binits{N.}},
\bauthor{\bsnm{Kovalev}, \binits{S.}},
\bauthor{\bsnm{Polley}, \binits{D.}},
\bauthor{\bsnm{Hagstr\"om}, \binits{N.Z.}},
\bauthor{\bsnm{Arekapudi}, \binits{S.S.P.K.}},
\bauthor{\bsnm{Semisalova}, \binits{A.}},
\bauthor{\bsnm{Lenz}, \binits{K.}},
\bauthor{\bsnm{Green}, \binits{B.}},
\bauthor{\bsnm{Deinert}, \binits{J.-C.}},
\bauthor{\bsnm{Ilyakov}, \binits{I.}},
\bauthor{\bsnm{Chen}, \binits{M.}},
\bauthor{\bsnm{Bawatna}, \binits{M.}},
\bauthor{\bsnm{Scalera}, \binits{V.}},
\bauthor{\bsnm{d'Aquino}, \binits{M.}},
\bauthor{\bsnm{Serpico}, \binits{C.}},
\bauthor{\bsnm{Hellwig}, \binits{O.}},
\bauthor{\bsnm{Wegrowe}, \binits{J.-E.}},
\bauthor{\bsnm{Gensch}, \binits{M.}},
\bauthor{\bsnm{Bonetti}, \binits{S.}}:
\batitle{Inertial spin dynamics in ferromagnets}.
\bjtitle{Nat. Phys.}
\bvolume{17},
\bfpage{245}
(\byear{2021})
\doiurl{10.1038/s41567-020-01040-y}
\end{barticle}
\endbibitem

\bibitem[\protect\citeauthoryear{Unikandanunni et~al.}{2022}]{unika22}
\begin{barticle}
\bauthor{\bsnm{Unikandanunni}, \binits{V.}},
\bauthor{\bsnm{Medapalli}, \binits{R.}},
\bauthor{\bsnm{Asa}, \binits{M.}},
\bauthor{\bsnm{Albisetti}, \binits{E.}},
\bauthor{\bsnm{Petti}, \binits{D.}},
\bauthor{\bsnm{Bertacco}, \binits{R.}},
\bauthor{\bsnm{Fullerton}, \binits{E.E.}},
\bauthor{\bsnm{Bonetti}, \binits{S.}}:
\batitle{Inertial {Spin Dynamics in Epitaxial Cobalt Films}}.
\bjtitle{Physical Review Letters}
\bvolume{129},
\bfpage{237201}
(\byear{2022})
\doiurl{10.1103/PhysRevLett.129.237201}
\end{barticle}
\endbibitem

\bibitem[\protect\citeauthoryear{De et~al.}{2023}]{de23}
\begin{botherref}
\oauthor{\bsnm{De}, \binits{A.}},
\oauthor{\bsnm{Lentfert}, \binits{A.}},
\oauthor{\bsnm{Scheuer}, \binits{L.}},
\oauthor{\bsnm{Stadtm\"uller}, \binits{B.}},
\oauthor{\bsnm{Pirro}, \binits{P.}},
\oauthor{\bsnm{Freymann}, \binits{G.}},
\oauthor{\bsnm{Aeschlimann}, \binits{M.}}:
Spin {Dynamics with Inertia in Ferromagnetic Thin Films}.
2023 IEEE International Magnetic Conference - Short Papers (INTERMAG Short
  Papers),
1
(2023)
\doiurl{10.1109/INTERMAGShortPapers58606.2023.10228822}
\end{botherref}
\endbibitem

\bibitem[\protect\citeauthoryear{De et~al.}{2024}]{de24}
\begin{barticle}
\bauthor{\bsnm{De}, \binits{A.}},
\bauthor{\bsnm{Schlegel}, \binits{J.}},
\bauthor{\bsnm{Lentfert}, \binits{A.}},
\bauthor{\bsnm{Scheuer}, \binits{L.}},
\bauthor{\bsnm{Stadtm\"uller}, \binits{B.}},
\bauthor{\bsnm{Pirro}, \binits{P.}},
\bauthor{\bsnm{Freymann}, \binits{G.}},
\bauthor{\bsnm{Nowak}, \binits{U.}},
\bauthor{\bsnm{Aeschlimann}, \binits{M.}}:
\batitle{Magnetic nutation: {T}ransient separation of magnetization from its
  angular momentum}.
\bjtitle{Physical Review B}
\bvolume{111},
\bfpage{014432}
(\byear{2024})
\doiurl{10.1103/PhysRevB.111.014432}
\end{barticle}
\endbibitem

\bibitem[\protect\citeauthoryear{Uhrig}{2025}]{uhrig25}
\begin{barticle}
\bauthor{\bsnm{Uhrig}, \binits{G.S.}}:
\batitle{Landau-{Lifshitz damping from Lindbladian dissipation in quantum
  magnets}}.
\bjtitle{New Journal of Physics}
\bvolume{27},
\bfpage{103502}
(\byear{2025})
\doiurl{10.1088/1367-2630/ae115c}
\end{barticle}
\endbibitem

\bibitem[\protect\citeauthoryear{Cygorek and Axt}{2014}]{cygor14}
\begin{barticle}
\bauthor{\bsnm{Cygorek}, \binits{M.}},
\bauthor{\bsnm{Axt}, \binits{V.M.}}:
\batitle{Comparison between a quantum kinetic theory of spin transfer dynamics
  in {Mn-doped bulk semiconductors and its Markov limit for nonzero Mn}
  magnetization}.
\bjtitle{Physical Review B}
\bvolume{90},
\bfpage{035206}
(\byear{2014})
\doiurl{10.1103/PhysRevB.90.035206}
\end{barticle}
\endbibitem

\bibitem[\protect\citeauthoryear{Stre\v{c}ka1 and
  Ja\v{s}\v{c}ur}{2015}]{strec15}
\begin{barticle}
\bauthor{\bsnm{Stre\v{c}ka1}, \binits{J.}},
\bauthor{\bsnm{Ja\v{s}\v{c}ur}, \binits{M.}}:
\batitle{{A BRIEF ACCOUNT OF THE ISING AND ISING-LIKE MODELS: MEAN-FIELD,
  EFFECTIVE-FIELD AND EXACT RESULTS}}.
\bjtitle{acta phys. slova.}
\bvolume{65},
\bfpage{235}
(\byear{2015})
\doiurl{10.48550/arXiv.1511.03031}
\end{barticle}
\endbibitem

\bibitem[\protect\citeauthoryear{Dijith et~al.}{2018}]{dijit18}
\begin{barticle}
\bauthor{\bsnm{Dijith}, \binits{K.S.}},
\bauthor{\bsnm{Aiswarya}, \binits{R.}},
\bauthor{\bsnm{Praveen}, \binits{M.}},
\bauthor{\bsnm{Pillai}, \binits{S.}},
\bauthor{\bsnm{Surendran}, \binits{K.P.}}:
\batitle{Polyol derived {Ni and NiFe} alloys for effective shielding of
  electromagnetic interference}.
\bjtitle{Mater. Chem. Front.}
\bvolume{2},
\bfpage{1829}
(\byear{2018})
\doiurl{10.1039/c8qm00264a}
\end{barticle}
\endbibitem

\bibitem[\protect\citeauthoryear{Zhang et~al.}{2019}]{zhang19c}
\begin{barticle}
\bauthor{\bsnm{Zhang}, \binits{X.}},
\bauthor{\bsnm{Lao}, \binits{Y.}},
\bauthor{\bsnm{Sklenar}, \binits{J.}},
\bauthor{\bsnm{Bingham}, \binits{N.S.}},
\bauthor{\bsnm{Batley}, \binits{J.T.}},
\bauthor{\bsnm{Watts}, \binits{J.D.}},
\bauthor{\bsnm{Nisoli}, \binits{C.}},
\bauthor{\bsnm{Leighton}, \binits{C.}},
\bauthor{\bsnm{Schiffer}, \binits{P.}}:
\batitle{Understanding thermal annealing of artificial spin ice}.
\bjtitle{APL Materials}
\bvolume{7},
\bfpage{111112}
(\byear{2019})
\doiurl{10.1063/1.5126713}
\end{barticle}
\endbibitem

\bibitem[\protect\citeauthoryear{Mukhopadhyay et~al.}{2023}]{mukho23}
\begin{barticle}
\bauthor{\bsnm{Mukhopadhyay}, \binits{S.}},
\bauthor{\bsnm{Majumder}, \binits{S.}},
\bauthor{\bsnm{Panda}, \binits{S.N.}},
\bauthor{\bsnm{Barman}, \binits{A.}}:
\batitle{Investigation of ultrafast demagnetization and gilbert damping and
  their correlation in different ferromagnetic thin films grown under identical
  conditions}.
\bjtitle{Nanotechnology}
\bvolume{134},
\bfpage{235702}
(\byear{2023})
\doiurl{10.1088/1361-6528/acc079}
\end{barticle}
\endbibitem

\bibitem[\protect\citeauthoryear{Jang et~al.}{2020}]{jang20}
\begin{barticle}
\bauthor{\bsnm{Jang}, \binits{H.}},
\bauthor{\bsnm{Marnitz}, \binits{L.}},
\bauthor{\bsnm{Huebner}, \binits{T.}},
\bauthor{\bsnm{Kimling}, \binits{J.}},
\bauthor{\bsnm{Kuschel}, \binits{T.}},
\bauthor{\bsnm{Cahill}, \binits{D.G.}}:
\batitle{Thermal {Conductivity of Oxide Tunnel Barriers in Magnetic Tunnel
  JunctionsMeasured by Ultrafast Thermoreflectance and Magneto-Optic Kerr
  Effect Thermometry}}.
\bjtitle{Phys. Rev. Appl.}
\bvolume{13},
\bfpage{024007}
(\byear{2020})
\doiurl{10.1103/PhysRevApplied.13.024007}
\end{barticle}
\endbibitem

\bibitem[\protect\citeauthoryear{Qader et~al.}{2017}]{qader17}
\begin{barticle}
\bauthor{\bsnm{Qader}, \binits{M.A.}},
\bauthor{\bsnm{Vishina}, \binits{A.}},
\bauthor{\bsnm{Yu}, \binits{L.}},
\bauthor{\bsnm{Garcia}, \binits{C.}},
\bauthor{\bsnm{Singh}, \binits{R.K.}},
\bauthor{\bsnm{Rizzo}, \binits{N.D.}},
\bauthor{\bsnm{Huang}, \binits{M.}},
\bauthor{\bsnm{Chamberlin}, \binits{R.}},
\bauthor{\bsnm{Belashchenko}, \binits{K.D.}},
\bauthor{\bsnm{Schilfgaarde}, \binits{M.}},
\bauthor{\bsnm{Newmana}, \binits{N.}}:
\batitle{The magnetic, electrical and structural properties of copper-permalloy
  alloys}.
\bjtitle{J. Mag. Mag. Mat.}
\bvolume{442},
\bfpage{45}
(\byear{2017})
\doiurl{10.1016/j.jmmm.2017.06.081}
\end{barticle}
\endbibitem

\end{thebibliography}



\end{document}